\documentclass{article}
\PassOptionsToPackage{numbers, compress}{natbib}
\usepackage[preprint]{neurips_data_2021}
\usepackage{makecell, cellspace, caption}

\usepackage{wrapfig,lipsum,booktabs}
\usepackage{amsmath,amssymb,amsfonts}
\usepackage{algorithmic}
\usepackage{graphicx}
\usepackage{textcomp}
\usepackage{wrapfig}
\usepackage{tcolorbox}

\usepackage{hyperref}
\hypersetup{
    colorlinks=true,
    linkcolor=blue,
    filecolor=magenta,      
    urlcolor=black,
    citecolor=black
}
\usepackage{lipsum}% http://ctan.org/pkg/lipsum
\def\BibTeX{{\rm B\kern-.05em{\sc i\kern-.025em b}\kern-.08emT\kern-.1667em\lower.7ex\hbox{E}\kern-.125emX}}
\definecolor{Gray}{gray}{0.3}
\usepackage[utf8]{inputenc} % allow utf-8 input
\usepackage[T1]{fontenc}    % use 8-bit T1 fonts
\usepackage{hyperref}       % hyperlinks
\usepackage{url}            % simple URL typesetting
\usepackage{booktabs}       % professional-quality tables
\usepackage{amsfonts}       % blackboard math symbols
\usepackage{nicefrac}       % compact symbols for 1/2, etc.
\usepackage{microtype}      % microtypography
\usepackage{xcolor}         % colors

\title{PyTorrent: A Python Library Corpus for Large-scale Language Models}

\author{Mehdi Bahrami$^{1}$
\And
N.C. Shrikanth$^{2}$
\And
Shade Ruangwan$^{3}$ 
\And
Lei Liu$^{1}$
\And
Yuji Mizobuchi$^{3}$
\And
Masahiro Fukuyori$^{3}$
\And
Wei-Peng Chen$^{1}$
\And
Kazuki Munakata$^{3}$ 
\And
Tim Menzies$^{2}$
\\ \\
$^{1}$Fujitsu Research of America, Sunnyvale, CA, USA
\\
$^{2}$North Carolina State University, Raleigh, NC, USA
\\
$^{3}$Fujitsu Research Ltd., Kawasaki, Japan
\\ \\
\texttt{\{mbahrami,ruangwan.shade,mizobuchi.yuji\}@fujitsu.com}
\\
\texttt{\{lliu,wchen,fukuyori,munakata.kazuki\}@fujitsu.com}
\\
\texttt{snaraya7@ncsu.edu timm@ieee.org}
}

\begin{document}
\maketitle
\begin{abstract}
% Code retrieval is allowing software engineers to search codes through a natural language query, which relies on both  natural language processing and software engineering techniques. Each code retrieval trained model requires
% There are some Python datasets that can be used to train a language model for Python machine-programming, such as code retrieval, code summarization, and code generation.
% All existing datasets for Python programming language focus on GitHub projects and relevant sources, such as forum discussions. However, in order to train a Python machine-programming model efficiently, we mainly need to address existing package libraries and their metadata as a first step that has been ignored in existing datasets. Importantly most of the Python packages are high-quality and well-documented.
% In this paper, we introduce PyTorrent, which is a large collection of 218,814 Python package libraries that include both metadata and harvested source-codes. PyTorrent can be used toward training and fine-tuning any downstream tasks of machine-programming.
% PyTorrent enables users (such as data scientists, students, etc.,) to build off the shelf machine learning models directly without spending months of effort on large infrastructure.
A large scale collection of both semantic and natural language resources is essential to leverage active Software Engineering research areas such as code reuse and code comprehensibility. Existing machine learning models ingest data from Open Source repositories (like GitHub projects) and forum discussions (like Stackoverflow.com),  whereas, in this showcase, we took a step backward to orchestrate a corpus titled PyTorrent that contains 218,814 Python package libraries from PyPI and Anaconda environment. This is because earlier studies have shown that much of the code is redundant and Python packages from these environments are better in quality and are well-documented. PyTorrent enables users (such as data scientists, students, etc.) to build off the shelf machine learning models directly without spending months of effort on large infrastructure. The dataset, schema and a pretrained language model is available at: \url{https://github.com/fla-sil/PyTorrent}

\end{abstract}
% \begin{IEEEkeywords}
% Python, machine-programming, code corpus;
% \end{IEEEkeywords}

\section{Introduction}
Presently Python is one of the most  popular~\cite{spectrum} programming language (as seen in  GitHub, StackOverflow, etc.). Numerous lines of codes are churned and accumulated in version control systems every day. 
Thus, if we can harness and organize `documented' Python code accumulated over these years, we could better attempt active Software Engineering (SE) problems in the lines of code reuse and code comprehensibility. 

At ICSE 2012, Hindle et al. \cite{hindle2012naturalness} conjectured that 
``\textit{..most software is also natural, in the sense that it is created by humans at work, with all the attendant constraints and limitations - and thus, like natural language, it is also likely to be repetitive and predictable}''. However, re-usability is beyond just the availability of data but the arduous task of searching through numerous artifacts to find the most relevant one for the target user (For example, Code Search~\cite{gu2018deep}). 

Understanding of a programming language relies on two pillars of source-code mining: i) the source-code of existing libraries and ii) the usage source-code of the existing libraries. Often researchers follow standard natural language techniques to generate a language model for programming languages (e.g., by processing a sheer number of library usages). However, programming languages are more restricted than natural languages (e.g., English), and the majority of code usage follows existing libraries. For instance, a model can learn about PyTorch~\cite{paszke2019pytorch} package from processing a sheer number of PyTorch code usages (existing approaches, i.e., mining GitHb projects that imported PyTorch) or mining source-code of the PyTorch package (proposed dataset). 

To that end we orchestrate a data set titled `\textit{PyTorrent}' that is made available public here~\cite{pytorrent-zenodo},\cite{pytorrent_github} and~\cite{pytorrent-v1-model}\footnote{PyTorrent Dataset: \url{https://doi.org/10.5281/zenodo.4451357}}.

A combination approach of mining both libraries and their usage, may improve the result with more precise outputs. In particular, Python programming language mainly relies on existing libraries where it allows users to easily install (i.e., using \textit{"pip install package"} for PyPI packages and \textit{"anaconda install package"} for Anaconda). 

\textbf{Background and Related Work.}
Several datasets for Python programming language have been released in recent years by processing GitHub repositories \cite{orru2015curated, biswas2019boa, markovtsev2018public, husain2019codesearchnet, agashe2019juice, clement2020pymt5}. 
The early works focus on particular use cases, and the dataset is relatively small. In \cite{orru2015curated}, a dataset of metrics taken from a curated collection on 51 popular software systems in Python has been published. The internal structure of each system is investigated, and metadata is collected (i.e., , including system identifier, description, etc. In another study, Biswas et al. \cite{biswas2019boa} create a dataset
that includes 1,558 mature GitHub projects for data science tasks. The Abstract Syntax Tree (AST) of parsed Python programs from each
revision along with the metadata are stored in the dataset. More recently, larger datasets have been generated. In \cite{markovtsev2018public}, the Public Git Archive has been released.
%The archive is the first big code dataset amenable to programmatic analysis. 
%It includes 182,014 top-bookmarked repositories written in different programming languages. 
% The dataset includes the URL of the Github repository, sova files which contains codes of that repo, as well as meta-data such as the programming language, commit count, branches count, fork count, number of lines and license. 
In \cite{husain2019codesearchnet}, the CodeSearchNet collects a corpus from publicly available open-source non-forked GitHub repositories and generates pairs of (comment, code). 
The schema of the corpus includes a limited number of fields of repository name, path of repository, function name, code, code tokens, docstring, docstring tokens, and programming language etc.  
The JUICE dataset presented in \cite{agashe2019juice} which collects publicly available Jupyter notebooks from GitHub and filter the notebooks having natural language markdown in English and Python as their kernel type. The JUICE dataset is composed of target cells that include at most one method definition with syntactically valid Python codes. 
In \cite{clement2020pymt5}, the PyMT5 dataset is presented by collecting 112K GitHub repositories. This method-docstring corpus includes the file level AST for each Python file, extracting every individual and class method. 

\par Although there are several datasets for Python programming language available, as aforementioned, these datasets mainly focus on GitHub projects and ignore the existing package libraries and their metadata (i.e., package name, package license, supported Python version, and package descriptions), which is one of the differences to our proposed PyTorrent dataset. Table \ref{tbl:comparision} summarizes the quantity comparison between our dataset and previous datasets. 
We focus on generating a dataset of the pair of natural language description (\textit{NL}) and functions/methods (\textit{PL}), for existing Python libraries. The pairs can be used for fine-tuning downstream tasks, such as code retrieval~\cite{gu2018deep}, code generation~\cite{yin2017syntactic} tasks and etc.
\begin{table}[t]
\centering
\footnotesize
\caption{Comparison of existing datasets}
\begin{tabular}{|l|l|l|l|l|}
\hline
\label{tbl:comparision}
%1\rowcolor{Gainsboro!60}
\textbf{Dataset} & \textbf{\# of Files} & \textbf{Lines of Codes} & \textbf{\# of Functions}\\ \hline
Curated Python \cite{orru2015curated} & Unknown & Unknown & Unknown\\ \hline
BOA \cite{biswas2019boa} & 4,977,680 & Unknown & Unknown\\ \hline
Public Git Archive \cite{markovtsev2018public} & 54.5 x $10^6$ & 15,941 x $10^6$ & Unknown\\ \hline
CodeSearchNet \cite{husain2019codesearchnet} & Unknown & Unknown & 2,273,157 \\ \hline
JUICE \cite{agashe2019juice} & 659,000 & Unknown & 1,521,774\\ \hline
PyMT5 \cite{clement2020pymt5} & 2.3 x $10^6$ & Unknown & 2.6 x $10^7$\\ \hline
%\rowcolor{Gainsboro!20}
\textbf{PyTorrent} (proposed) & 4,058,349 & 655,074,611 & 2,841,446\\ \hline
\end{tabular}
\end{table}

\section{Methodology}
\begin{figure}[t]
    \centering
    \includegraphics[width=11cm]{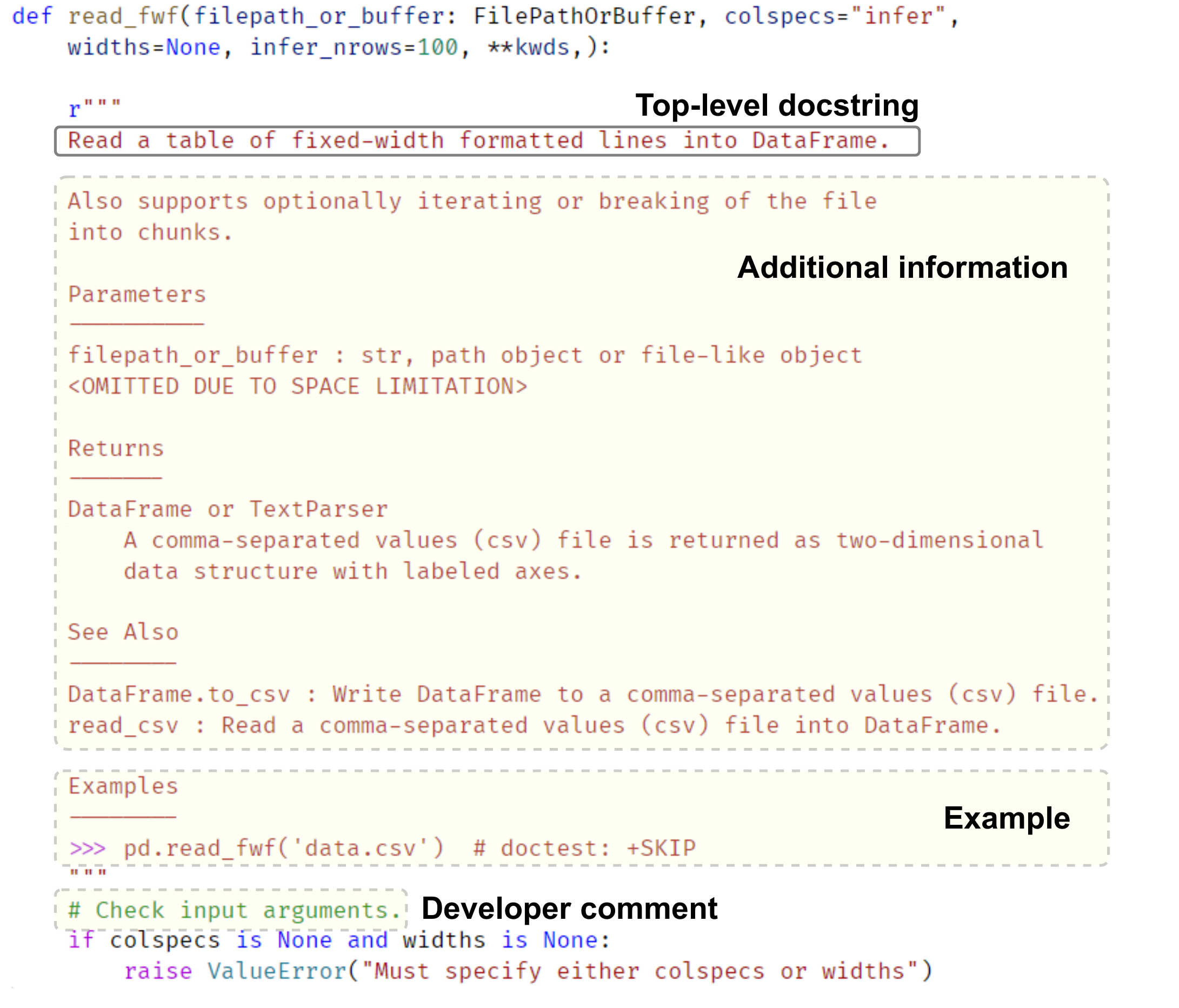}
    \caption{Snapshot of a well documented method in pandas library}\footnote{https://pandas.pydata.org/}
    \label{fig:entities}
\end{figure}

\textbf{Dataset Description.} Our goal was to mine voluminous \textit{\textless PL,NL\textgreater} (Source code, Natural language text) pairs (see Figure~\ref{fig:entities}) to promulgate SE research. Our focus was to orchestrate a corpus that is compliant with existing datasets for machine programming language models and, in particular, transformer based models~\cite{wolf2019huggingface}. Therefore, we choose the schema of CodeSearchNet dataset~\cite{codesearchnet_format} as base schema and generates PyTorrent dataset according to the defined schema. 

\par However, we differ from CodeSearchNet in three ways. Firstly, Unlike CodeSearchNet, our data source is from Python libraries from PyPI and Anaconda packages rather than GitHub projects. 
\par Secondly, CodeSearchNet builds machine learning models only based on the top-level docstring description. In our case, we generate three augmented datasets: i) top-level docstring, ii) added user comments, and iii) added both full-docstring and user comments where we associate Python methods with the entire docstring description and extract developer comments as marked in Figure~\ref{fig:entities}.
    
\par Lastly,  the two \textit{primary} attributes in CodeSearchNet schema are \textit{docstring\_tokens} and \textit{code\_tokens}. The \textit{docstring\_tokens} act as a placeholder for natural text where \textit{developer comments} and (or)  \textit{docstrings} can be tokenized and assigned, correspondingly tokenized source code  can be assigned to \textit{code\_tokens}. Then one can build a machine learning model using the aforementioned architectures like CodeSearchNet. Some attributes like repo, path, and URL are set to Python package name, Python script path. For the first time, researchers may use package name in the PyTorrent dataset to define a cross join relation between pairs and the corresponding package metadata. We also added \textit{docstring\_summary}. The detail schema can be found in PyTorrent GitHub repository ~\cite{pytorrent_github}.
    
% \end{itemize}

\textbf{Data Extraction.}
The first step of producing PyTorrent is collecting metadata of existing libraries from Python package distributions that include PyPI~\cite{pypi} and Anaconda~\cite{anaconda}. Each Python package distribution offers different metadata of each software package.
We utilize Scrapy~\cite{scrapy} to crawl packages through PyPI API and Anaconda API. Our crawler collects and generates an index of Python library packages for PyTorrent. The native API offers only raw text version of metadata. Therefore, we use Scrapy to collect enriched HTML package descriptions. We published the final updated schema and all packages that include PyPI package metadata schema \cite{pypi_schema}, Anaconda package metadata schema and all  harvested metadata of packages~\cite{pytorrent_github}.
We construct 238,187 package  metadata~\cite{pytorrent_github}.
Once we collect the package metadata, we collect the latest version source-code of each package to construct PyTorrent. The web crawler collected 218,814 packages with source-codes.

The collected decompressed raw files include 10,449,240 files for all Python software packages as of December 2020. The raw collection size is 100+ GB of compressed data and 379 GB of decompressed (including 62,570 different file extension) of raw data which includes 4,058,349 Python scripts with the size of 30 GB.
Each package of PyTorrent in metadata includes the number of releases. We use a minimum number of releases as a threshold to list the number of packages in each category, which is shown in Table \ref{tbl:releases}. As the statistical shown, a large number of Python packages (4,296) with the highest updates (i.e., threshold=50) may indicate the most popular and well-documented packages. For generating PyTorrent, we consider only the latest package release source-codes.
\begin{table}[t]
\footnotesize
\centering
\caption{The Number of packages per minimum number of releases}
\begin{tabular}{|c|r|c|}
\hline
\label{tbl:releases}
%\rowcolor{Gainsboro!60}
\textbf{Threshold} & \textbf{\# of Packages} & \textbf{Compressed Size } \\ \hline
%\rowcolor{Gainsboro!20}
1 & 218,814 & 100.8 GB (\textbf{379 GB raw}) \\ \hline
2 & 129,710 & 74.85 GB \\ \hline
5 & 76,267 & 55.88 GB \\ \hline
10 & 40,692 & 47.92 GB \\ \hline
50 & 4,296 & 13.26 GB \\ \hline
\end{tabular}
\end{table}

\textbf{Data Pre-processing.}
%\textbf{Data Source.}
% Why Python
%Most studies in this space~\cite{husain2019codesearchnet,feng2020codebert} only consider the `Top-level docstring' as shown in Figure~\ref{fig:entities}, whereas as shown in that figure, there is a rich source of untouched information that we believe can leverage machine learning models. 
A typical example of a well-documented method in a library is shown in Figure~\ref{fig:entities}. While most studies such as ~\cite{husain2019codesearchnet,feng2020codebert} only consider the `Top-level docstring',  we believe the rich source of additional information could be utilized to strengthen machine learning models. 
Information beyond just top-level docstring is mostly available as part of any well-documented code. Typically it includes information about exceptions, return types, and most important, usage information (example). Further, the method body would include developer comments describing the purpose of the code written in natural language succinctly. Note that although the docstring is written in the natural language, it is contained within some templates. Docstring can be written in numerous styles like Google, NumPy/SciPy, reStructured Text, and Epytext. On the other hand, developer comments are unstructured similar to google search queries that may not adhere to a syntax (like Structured Query Language). We presume it would help machine learning models easily retrieve code when trained with such natural language (developer comments) artifacts, especially when the input is a similar natural language search query.

\par Each package is decompressed and process each script of packages. We use all packages with $Threshold=1$ to generate the final dataset. 

Each package may include a set of Python scripts and other resources (i.e., test cases, readme file). We process only Python scripts.\\

\textbf{Dataset Construction.} In a nutshell, all core Python scripts (as defined in Table~\ref{tbl:code_category}) are parsed and stored in a JSONL~\cite{jsonl} format. First, we extract all functions (both class level and script level) and their associated docstrings. Then, we extract developer comments and their associated source codes within the scripts (including lines of code within each function). 
 
We have three objectives here (refer to Figure~\ref{fig:entities}):
\begin{itemize}
    \item To store a Python class level function or script level function and its associated docstring.
    \item To store source-code with its associated developer comments (as part of \textit{PL} at every level), such as:
    \begin{itemize}
        \item Code that is part of the script (not bound by any class or function)
        \item Code that is within a function (class or script level function)
        \item The function itself (class or script level)
    \end{itemize}
\end{itemize}
To achieve this, we take advantage of the Abstract Syntax Tree (AST) and source-code line number information to associate the appropriate developer comments or docstrings with its associated source codes. For each Python script, we do the following.
\begin{itemize}
    \item Visit each Python function (both script~\footnote{Python function part of the script but does not belong to any class} level and class~\footnote{Python function that is part of a class within that script} level) and store its raw source code and its docstring. Then, we visit each of these functions to identify a line that begins with \# (to identify developer comment). Next, we associate all source code under that developer comment part of the method.
    
    \item Using source code line number information, we ignore all the functions that are processed earlier and consider the rest of the code in the Python script to do the following:
    i)~We identify a line that begins with \# (to identify developer comment);
    ii)~Next, we associate all source code under that developer comment part of the method.
\end{itemize}

All the pairs parsed from the Python scripts are written to disk into a JSON format as they are visited in each of the packages. To make dataset compliance with other related language model datasets for Python, we saved contents as JSON Lines text format~\cite{jsonl} and split into three folders of \textit{valid, test} (284,145, 284,144 pairs) and \textit{train}(2,273,157 pairs), the pairs in each folder split into 20 chunks and compressed through gzip. In order to save dataset as a single file for archiving dataset, we use gzip. We generate three compressed file for three different scenarios of \textit{UserComments}, added \textit{Docstrings} and \textit{both} docstrings and user comments. The size of each compressed file is \~4.1~GB~\cite{pytorrent-zenodo} and the size of each decompressed (JSONL) file is 43~GB\cite{pytorrent-zenodo}~\cite{pytorrent_github}.
%%%%%%%%%%%%%%% ADDED %%%%%%%%%%%%%%

We constructed models using five scenarios listed in Table \ref{tbl:scenarios}. 
%%%%%%%%%%%%%%%%%%%%%%%%%%%%%%%%%%% TODO: MOVE to CONTEXT %%%%%%%%%%%%%%%%%%%%%%%%%%%%%%%%%%%% 
\begin{table}[ht]
\centering
\caption{5 Augmented-Code Scenarios (ACS) to build code retrieval models ($X  \Longrightarrow Y$) in CodeSearchNet and CodeBERT architectures (ACS=0 is the default case for CodeSearchNet and CodeBERT, whereas our results endorse ACS=4).  }
\begin{tabular}{|p{2cm}|p{5cm}|p{5cm}|}
\hline
\label{tbl:scenarios}
\textbf{ACS} & \textbf{X} (attribute:$docstring\_tokens$)  & \textbf{Y} (attribute:$code\_tokens$) \\
\hline
\textbf{0 (default)} & Tokenized short description of docstring  & Tokenized (code and code comments) \\ \hline
1                 & Tokenized code comments             &     Tokenized code                       \\

2                 & Tokenized (code comments  and entire docstring)                                                & Tokenized code                                                   \\

3                 & Tokenized (code comments, entire docstring and commit message)                         & Tokenized (code and  code comments) \\  

4                 & Tokenized (code comments  and entire docstring)    & Tokenized (code and code comments)                    \\

5                 &  Tokenized short description of docstring                 & Tokenized code                             \\ \hline
\end{tabular}

\end{table}
%%%%%%%%%%%%%%%%%%%%%%%%%%%%%%%%%%% END TODO %%%%%%%%%%%%%%%%%%%%%%%%%%%%%%%%%%%% 
\textbf{Data filtering.} Each extracted pairs of \textit{\textless PL,NL\textgreater} can be categorized in one of the four listed in Table %
\ref{tbl:code_category}.
\begin{table}%{r}{4.5cm}
\centering
\footnotesize
\caption{Number of pairs per source-code category}
\begin{tabular}{|l|l|}
\hline
\label{tbl:code_category}
Code Category & \# of Pairs \\ \hline
Core & 11,324,635 \\ \hline
Test & 1,632,686 \\ \hline
Init & 635,108 \\ \hline
Other & 233,218 \\ \hline
\end{tabular}
\end{table}
The table shows all categories of source-code and their number of pairs. We came up with these categories after inspecting a small portion of the extracted Python scripts. This categorization primarily helps filter noisy scripts in our dataset (files like \_\_init\_\_.py) that do not contain reusable code. Secondly, researchers may build focused machine learning models. For example, to write test cases, one may create a Code Search machine model only using scripts part of the \textit{Test} (identified if the script path contains any case of test) category.
Scripts specifically such as \textit{setup.py} and \textit{make.py} are categorized as \textit{Other}. 
The remaining Python scripts (82\%) do not fit in the three categories is placed in the core category. PyTorrent is orchestrated based on the selected scripts in the \textit{Core} category that has content of function and docstrings with a total of 2,841,446 pairs.

\textbf{Data Insights.}
\label{sec:Data Insights.}
To give a general idea of PyTorrent, we perform preliminary analysis on its source codes and docstrings.
In particular, we calculate code-related metrics from the collected source codes using \texttt{Radon} library~\cite{url_radon}.
We show the number of lines and words for docstrings. From selected 2,925,660 functions in PyTorrent, we find that on average, source codes have McCabe’s complexity\cite{mccabe1976complexity} of 3 with a median of 1.
Docstrings, on average, contains 6~lines (median~3) and 31~words (median 15).
% source codes have 10 (5) lines of code, 9 (5) logical lines of code
In addition, we find that 25\% of those functions use Python decorator.
Our observation of PyTorrent shows that the average comment length of all pair is 47 characters, an average code length is 228 characters, and 228,426 of pairs include examples (i.e., Tox-based~\cite{tox}). 
Table~\ref{tbl:Percentile_whole} shows the detail of a percentile of length of NL and PL separately for the whole dataset. 
%\begin{wraptable}{r}{3.5cm}
\begin{table}
% \begin{table}[t]
\centering
\footnotesize
\caption{Percentile length coverage per NL \& PL (whole PyTorrent)}
\begin{tabular}{|l|l|l|}
\hline
\label{tbl:Percentile_whole}
Percentile&NL& PL \\ \hline
0.50 & 9 & 44 \\ \hline
0.70 & 11 & 77 \\ \hline
0.80 & 12 & 105 \\ \hline
0.90 & 14 & 165 \\ \hline
0.95 & 16 & 243 \\ \hline
%ref: PBot_Mehdi_2020_07_28_v1.pptx
\end{tabular}
\end{table}

\begin{figure}%{r}{1.0\textwidth}
    \centering
    \includegraphics[width=6cm]{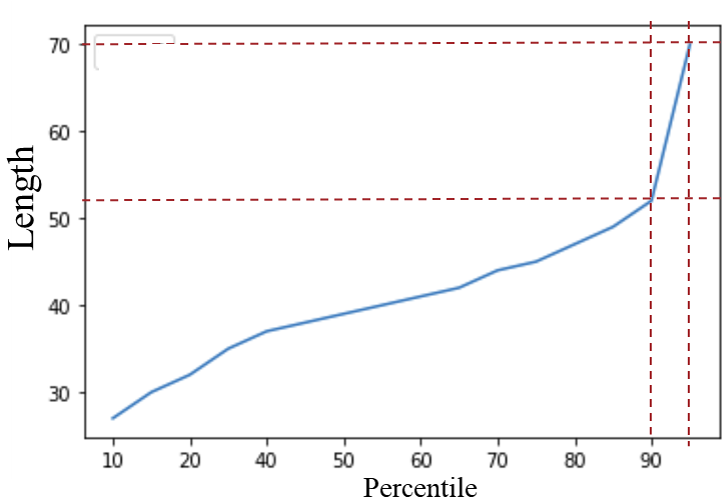}
    \caption{Percentile length of LoC for a coverage of 66,528 sample scripts}
    \label{fig:Percentile_coverage}
\end{figure}

Although the average length of Line of Code (LoC) is 47 in all PyTorrent packages, we investigate it with more precision on a portion of the PyTorrent dataset for 66,528 sample raw Python scripts. The sample data includes 9,951,811 LoC, and the average number of LoC in each script includes 149. In this sample dataset, we observe the length of LoC with an average of 44.92 (which is very close to the whole dataset with the length of 47). Interestingly, we observe that selecting a length of 50 and 70 covers 61,854 (\%86) and 65,656 (\%99) of all sample records without truncation, respectively. The detail of percentile coverage is shown in Figure~\ref{fig:Percentile_coverage}. Therefore, base on the statistic results \textit{we recommend that researchers safely truncate a LoC Python programming language with a much smaller length (i.e., 70)} in comparison to natural language models (i.e., 512). Selecting a smaller size allows the researcher to generate a smaller tokenizer and less LoC encoder size to save computation resources and much speedy processing on both fine-tuning and inferring from a BERT-based language model~\cite{devlin2018bert}.

\textbf{Source Code.}
Notably, we collected the data on two 2.20GHz Intel CPUs having 56 cores with 32 GB RAM. The artifacts of PyTorrent include the schema of raw packages, package metadata,~
\cite{pytorrent_github} a compliant pair of \textit{\textless NL,PL\textgreater} dataset~\cite{pytorrent-zenodo}, a preliminary transformer-based DistilBERT model~\cite{pytorrent-v1-model} have been published.

\section{PyTorrent: Why \& How ?}

\textbf{Why?} Most of the tedious work lies in preparing the data before insights can be mined~\cite{pyle1999data}. Any data showcase's primary objective is to enable users to readily build data models (similar to UCI Machine Learning Repositories datasets~\cite{asuncion2007uci}). There are two phases in the construction of the PyTorrent dataset. In the first phase, We collected all Python software packages that include 218,814 modules. In the second phase, we visited every Python script within each of the mined modules to extract Python methods, code snippets, and associated documentations. We spent 12 weeks of effort in web crawling, data extraction, and pre-processing.

%%%%%%%%%%%%%%%%%%%%%%%%%%%%%%%%%%%% TODO: move to context %%%%%%%%%%%%%%%%%%%%%%%%%%%%%%%%%%%% 
%
% % Please add the following required packages to your document preamble:
% % \usepackage[table,xcdraw]{xcolor}
% % If you use beamer only pass "xcolor=table" option, i.e. \documentclass[xcolor=table]{beamer}
% \begin{table}[]
% \centering
% \caption{In total TBA SLOC of code was written over 3 month period to extract, pre-process and transform TBA modules to build the PyTorrent dataset. }
% \label{tbl:effort}
% \begin{tabular}{|p{1.5cm}|p{1.5cm}|p{2cm}|p{2cm}|}
% \hline
% \rc \textbf{Package} & \textbf{\#Modules} & \textbf{Extraction Time (hours)} & \textbf{PyTorrent Construction (hours)} \\ \hline
% PyPi             &                     &                                  &                                        \\ \hline
% Anaconda         &                     &                                  &                                        \\ \hline
% \end{tabular}

% \end{table}
%%%%%%%%%%%%%%%%%%%%%%%%%%%%%%%%%%%% END TODO %%%%%%%%%%%%%%%%%%%%%%%%%%%%%%%%%%%% 
% \textbf{SE Research.}
% The data from PyTorrent can leverage research areas such as Code Search and Automated  Program Repair (but not limited to only these). 
% 
% 
% 
% 
%
%%%%%%%%%%%%%%%%%%%%%%%%%%%%%%%%%%%% TODO %%%%%%%%%%%%%%%%%%%%%%%%%%%%%%%%%%%% 
% \section{Who should use PyTorrent?}

% \begin{itemize}
% \item Researchers
% \item Practitioners
% \item Large enterprise, Students etc
% \end{itemize}
%%%%%%%%%%%%%%%%%%%%%%%%%%%%%%%%%%%% END TODO %%%%%%%%%%%%%%%%%%%%%%%%%%%%%%%%%%%% 

% \section{How to use PyTorrent with CodeSearchNet?}

\textbf{How?} This section describe steps to build a CodeSearchNet architecture model~\cite{husain2019codesearchnet} with PyTorrent dataset.

i)~Setup the CodeSearchNet~\cite{husain2019codesearchnet} environment or similar environment; ii)~The PyTorrent is a vast JSON file that split into 20 chunks; %Hence a simple script with few lines can be written to transform that JSON to a schema adhering to an underlying architecture (in this case, as JSON format required by CodeSearchNet);
iii)~dataset includes \textit{train} (80\%), \textit{test} (10\%), and \textit{valid} (10\%) folders and place them in the corresponding folder within the CodeSearchNet environment; iv)~One should be now able to build CodeSearch deep learning models with an issue of a simple command. For example, the following CodeSearchNet command $python$ $train.py$ $--model$ $neuralbow$ trains a Neural Bag of Words model based on the PyTorrent.

The steps above show how easily PyTorrent corpus can put to the test in practice (but not limited to CodeSearch).

\section{Potential Research Applications}

\textbf{Conjecture.} Code Search, Code Generation, Defect Prediction, and Program Repair are some of the prevalent techniques in the Software Engineering (SE) space aimed to accelerate developer productivity and software quality.  In Code search ~\cite{gu2018deep,husain2019codesearchnet,feng2020codebert} the problem is to accurately map a natural language query with a relevant source code; such a technique typically integrated into an IDE (e.g., IntelliSense~\cite{url_visualstudio}) to improve developer productivity (writing more lines of code in less time).
Code generation~\cite{gulwani2017program} is similar to code search problem, but instead of mapping a query with relevant source code, a language model generates source code according to a query with its own knowledge learned during the training process.
Defect prediction is a decades-old and active research space where the problem is to accurately predict defects~\cite{d2012evaluating,nam2017heterogeneous}. But post prediction, another functional research space, namely automated program repair~\cite{nguyen2013semfix,liu2018mining} attempt to patch appropriate fixes to reduce developer debugging time. PyTorrent contains voluminous source code information extracted from widely used modules (such as pandas~\cite{url_pydata}, and Pytorch) that we believe would accelerate the SE research.

We presume these widely used packages are well tested. Hence source code snippets from those packages can be suitable candidates to patch (repair) defective code (i.e., code that fails some test). 

% \textbf{Software Quality.}

\textbf{Usage.}
Since we generate a compliant dataset with a recent programming language models' dataset, PyTorrent can be easily plugin into CodeSearchNet~\cite{husain2019codesearchnet}, CodeBERT~\cite{feng2020codebert}, CodeXGLUE~\cite{CodeXGLUE} or GraphCodeBERT~\cite{ling2020deep} for training or fine-tuning a machine programming language models or fine-tuning on a downstream task. For instance, first download dataset from one of the augmented code datasets which is available here \cite{pytorrent-zenodo}, or \cite{pytorrent_github}. Then, i) to train a CodeSearchNet model, add downloaded PyTorrent dataset to \textit{`/resources/data'} and run training script as explained in \cite{CSNET_traing}; ii) to fine-tune a CodeBERT model, add downloaded PyTorrent dataset to \textit{`/data/codesearch`} and run fine-tuning script as explain in \cite{codebert_finetuning}. The similar approach can be used for fine-tuning CodeXGLUE and GraphCodeBERT.
%
%
%The generated PyTorrent dataset can be used for training or fine-tuning a language model for Python programming language. 
We use PyTorrent to train a preliminary DistilBERT-Masked Language Modeling(MLM)~\cite{sanh2019distilbert} model from scratch. The trained model, along with the dataset, aims to help researchers to easily and efficiently work on a large dataset of Python packages using only 5~lines of codes to load the transformer-based model. We use 1M raw Python scripts of PyTorrent that includes 12,350,000 LOC to train the model. We also train a byte-level Byte-pair encoding (BPE)~\cite{wang2016character} tokenizer that includes 56,000 tokens, which is truncated LOC with the length of 50 to save computation resources as explain in Section~\ref{sec:Data Insights.}. We published the transformer-based model at HuggingFace Hub as PyTorrent(v1)~\cite{pytorrent-v1-model}. 

Another application of PyTorrent dataset lies in code generation where developers input natural language description of code.
The model then generates an executable source code according to the description.
To train aforementioned model, many pairs of \textit{\textless NL,PL\textgreater} preferably high on quality are needed.
A code generation model can potentially support developers to shorten development time and help newcomers in learning from good examples.
To explore this possibility, we use PyTorrent to fine-tune the pre-trained CodeBERT~\cite{codebert_finetuning}.
The results are promising especially in multi-lines code generation. \\
\par The dataset can be used to produce an augmented programming language\cite{bahrami2021augcode} where it allows the dataset to be extended for fine-tuning a model on a downstream task, such as code generation, code search and etc.

\section{Conclusion}
In this paper, we introduce a new dataset of Python programming language (PyTorrent) that mined 218,814 Python library packages source-codes. The dataset includes different augmented code scenarios where it allows both practitioners and researchers to efficiently plugin the dataset for training and fine-tuning a language models. We published ii) metadata of all packages, ii) three augmented code scenarios, and iii) a pre-trained transformer model~\cite{pytorrent_huggingface}.
iii) a preliminary results on two downstream tasks of code retrieval and code generation tasks.

%%%%%%%%%%%%%%%% ADDED %%%%%%%%%%%%%%%%

\section{Limitations and Future Work}\label{sec:lim_fut}
Abdalkareem et al. \cite{abdalkareem2020impact} recently found that around 10\% to 16\% of the packages within the prevalent package managers are trivial. Trivial packages are largely preferred by software engineers, and their utility is observed in large organizations, including Facebook and Netflix. Abdalkareem et al. define a trivial package as a package with $\le 35$ LOC and McCabe’s cyclomatic complexity $\le 10$. Further, there are other package management platforms like \textit{npm} (JavaScript) with a large developer audience. Thus a natural extension of this work would be to mine \textit{npm} packages and analyze trivial packages in all 218,814 packages we mined.
%We also plan to publish results of downstream tasks such as code generation on transformer-based PyTorre
%We also plan to publish results of downstream tasks such as code generation on transformer-based PyTorrent model.

%\bibliographystyle{IEEEtran}
%\bibliographystyle{plainnat}
%\bibliography{main}
\end{document}